\documentclass{article}
\usepackage{frascatiphys}
\usepackage{graphicx}
\usepackage{hyperref}
\usepackage{amssymb}
\begin{document}
\title{ 
Photons in the science of the Pierre Auger Observatory
}
\author{
  Sergio Petrera$^{1,2}$ for the Pierre Auger Collaboration$^{3,4}$\\
  {$^1$ \em Gran Sasso Science Institute (GSSI), L'Aquila, Italy}\\
  {$^2$ \em INFN Laboratori Nazionali del Gran Sasso, Assergi (L'Aquila), Italy}\\ 
  {$^3$ \em Observatorio Pierre Auger, Av. San Mart\'in Norte 304, 5613 Malarg\"ue, Argentina}\\
  {$^4$ \em Full author list: \href{http://www.auger.org/archive/authors_2019_06.html}{\rm http://www.auger.org/archive/authors\_2019\_06.html}}
  }
\maketitle
\baselineskip=11.6pt
\begin{abstract}
  In this paper, the connection between Pierre Auger Observatory  measurements and photons
  is discussed. Three cases are presented: the search for photons in the ultrahigh-energy cosmic-ray radiation, the impact of the photon background in the cosmic ray propagation and the role of the ambient photon fields surrounding cosmic accelerators.
\end{abstract}
\baselineskip=14pt

\section{Introduction}
The Pierre Auger Observatory\cite{ThePierreAuger:2015rma} is the largest facility to detect cosmic rays built so far.
It is located in the province of Mendoza,
Argentina and has been in operation since 2004. Cosmic rays (CR) are studied by combining a 
Surface Detector (SD) and a Fluorescence Detector (FD) to measure extensive air showers.
The SD consists of 1600 water-Cherenkov detectors on a 1500 m triangular grid (SD-1500) over an area of about 3000 km$^2$ and additional 61 detectors covering 23.5 km$^2$ on a 750 m grid (SD-750 or `infill' array).
The 24 fluorescence telescopes grouped in four FD buildings are located on the boundary of the observatory to overlook the whole atmospheric volume above the surface array. Three additional telescopes pointing at higher elevations (HEAT) are located near one of the FD sites (Coihueco) to detect lower energy showers.
An array of radio antennas, Auger Engineering Radio Array (AERA)\cite{ThePierreAuger:2015rma,Abreu:2012pi}, complements the data with the detection of the shower radiation in the hundred MHz region.

The design of the observatory has been conceived to exploit the `hybrid' concept, the simultaneous detection of air showers by the surface array and fluorescence telescopes. The apparatus collects shower events of different classes depending on the on-time (generally called duty cycle) of the different detector components: the surface array is able to collect showers at any time, whereas the fluorescence detectors can operate only during clear moonless nights ($\approx$ 15\% duty cycle).  After taking into account geometry and quality cuts applied at the event reconstruction level, the hybrid data-set is only a few percent. Therefore only a small part of the SD showers are actually reconstructed by the FD. Nonetheless, this sub-sample (the hybrid data-set) is very valuable, including events having both the footprint of the shower at the ground and the longitudinal profile measured.  The hybrid approach has been a major breakthrough in the detection of UHE cosmic rays (UHE stands for $E > 10^{18}$ eV) since the method allows one to have the same energy scale in the surface detectors and the fluorescence telescopes and to derive the energy spectra entirely data-driven and free of model-dependent assumptions about hadronic interactions in air showers.

The science outcomes of the Pierre Auger Observatory are numerous and address several features of UHE cosmic rays, like the energy spectrum, the mass composition and the anisotropies in the arrival directions. It is beyond the purposes of this paper to show these results: the most recent ones can be found in ref.\cite{HL-ICRC2019}. In this paper, I discuss how photons are related to the Auger results both as CR particles and as background fields. In particular, in Sec. \ref{sec:Search}, I summarize the advances in the search of UHE photons. The existence of these photons and their fraction to CR nuclear particles give remarkable hints about their origin. On the other hand, photons also affect UHECR observables (e.g. spectrum and composition) because cosmic rays interact with the photon fields present in the sources and the Universe. In Sec. \ref{sec:PhBkgr}, I discuss the impact of the photon background in the UHECR propagation, which connects the cosmic-ray sources to the observables we measure. Finally, in Sec. \ref{sec:PhSources}, I discuss the influence of the photon fields surrounding the cosmic-rays sources on the same observables.

\section{Search for Ultrahigh-Energy Photons}
\label{sec:Search}

A flux of photons with energies above 1 EeV is expected from the decay of $\pi^0$ particles produced by protons interacting with the cosmic microwave background (CMB) in the so-called Greisen-Zatsepin-Kuz'min (GZK) effect. In this scenario, called `bottom-up', photons originate from the propagation of CR particles through the photon background and their flux is directly connected to the primary CR flux. The expected flux of GZK photons is estimated to be of the order of 0.01-0.1\% of the total CR flux, depending on the astrophysical model\cite{Gelmini:2005wu,Hooper:2010ze,SKK} (e.g., mass composition and spectral shape at the source). 
Instead, a large flux of UHE photons is predicted in `top-down' models: in this scenario\cite{Bhattacharjee:1998qc,Aloisio:2015lva}, ultrahigh-energy cosmic rays originate from the decay of supermassive particles, and these particles have decay branching ratios into photons (as well as neutrinos) comparable to that into hadrons. For this reason, UHE photon (and neutrino) limits are powerful tools to discriminate between the two scenarios. Further, it is worth noticing that a possible non-observation of UHE photons is meaningful for the physics foundations at the highest energies because it provides constraints to Lorentz Invariance Violation (LIV)\cite{Galaverni:2007tq}, QED non-linearities\cite{Maccione:2008iw} and space-time structures\cite{Maccione:2010sv}.

\begin{figure*}[b!]
    \centering
    \begin{tabular}{ll}
      \includegraphics[scale=0.26]{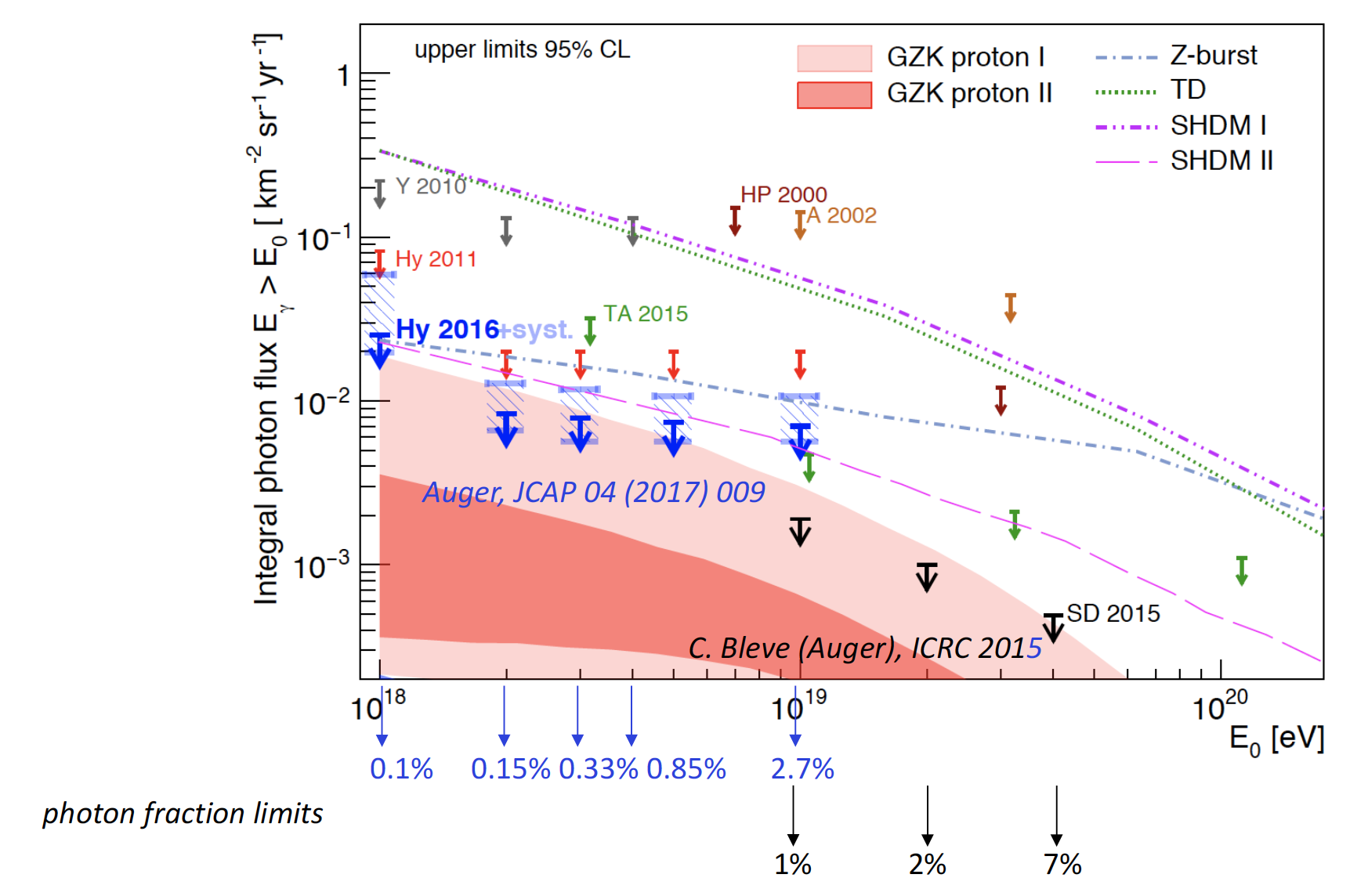}
      &
        \includegraphics[scale=0.55]{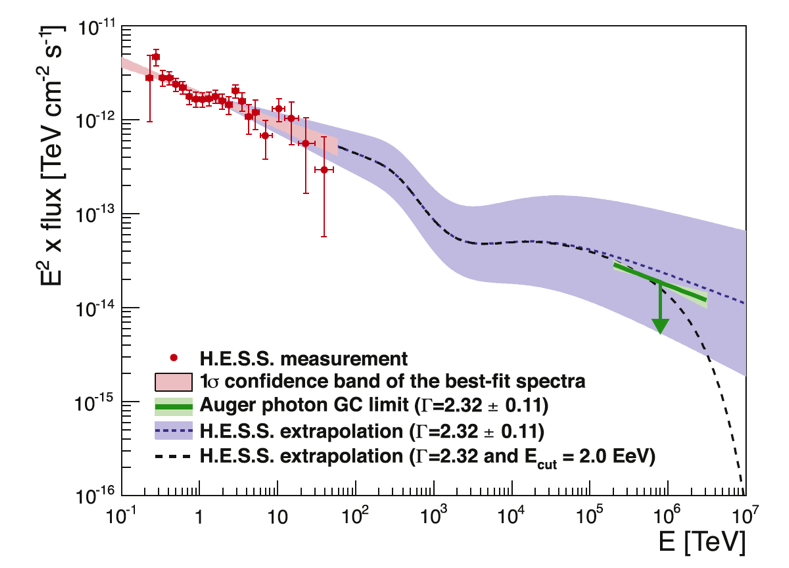}
        
    \end{tabular}
\caption{{\it Left: Upper limits on the integral photon flux from Auger for a photon flux $E^{-2}$ and no background subtraction. The corresponding photon fractions are also given. Other limits from Telescope Array (TA), AGASA (A), Yakutsk (Y) and Haverah Park (HP)  are shown for comparison. The
shaded regions and the lines give the predictions for the GZK photon flux  and for top-down
models. References to data and models in\cite{Aab:2016agp}.
Right: Photon flux as a function of energy from the Galactic center region.
Measured data by H.E.S.S. are indicated, as well as the extrapolated photon
flux at Earth in the EeV range. The Auger limit is
indicated by a green line. A variation of the assumed spectral index by $\pm$0.11
according to systematics of the H.E.S.S. measurement is denoted by the light
green and blue band. References in\cite{Aab:2016bpi}.}}
\label{fig:PhotonLimits}
\end{figure*}

In the Pierre Auger Observatory, UHE cosmic rays are studied by observing the extensive air showers (EAS) originating from their interactions with the atmosphere. Therefore, the nature of the primary is analysed looking at mass parameters which exhibit different sensitivities to photon and hadron showers. EAS initiated by UHE photons have two remarkable features: a delayed development of the shower profile and a reduced muon content. To give an idea of the separation induced by the shower development, simulated proton
and photon showers have average depths of the shower maximum, $X_\mathrm{max}$, that differ by about 200 g/cm$^2$ in the EeV ($10^{18}$ eV) range. The lower muon content is instead detectable at the ground using the SD, where smaller footprints, steeper lateral distributions and faster rise-times are expected for photons. 
The observables used in the photon searches are different depending on the
primary energy:
\begin{itemize}
\item At lower energies, hybrid events are numerous. For these events,
  the depth of the shower maximum, $X_\mathrm{max}$, is the most sensitive mass parameter. Searches based on severe cuts on the measured $X_\mathrm{max}$ 
  have been published\cite{Abraham:2006ar,Abraham:2009qb} where upper limits have been set for energies above 1, 2, 3, 5 and 10 EeV. The latest search based on the same event class mixes $X_\mathrm{max}$ with two SD observables, $S_\mathrm{b}$ and $N_\mathrm{stat}$\footnote{
    See ref.\cite{Aab:2016agp} for their definition.},  which show sensitivity  to the separation between photons and hadrons. A multivariate
    analysis based on the Boosted Decision Tree (BDT) technique has allowed us to improve the previous limits\cite{Aab:2016agp}.
  \item At higher energies only SD observables are used to search for photon signatures. Upper
    photon flux limits  for energies above $10^{19}$ eV have been published in \cite{Aglietta:2007yx,Bleve:2015nut}.  
  
  \end{itemize}
  In figure \ref{fig:PhotonLimits} left panel, the Auger results on the integral photon flux are shown in comparison with the results from other experiments and model predictions. The achieved sensitivity allows testing photon fractions of about 0.1\% at EeV energies and percent level at higher energies. This outcome rules-out early top-down models and challenges the most recent
  super-heavy dark matter models. Furthermore, these values initiate the
  exploration of the region of photon fluxes predicted in GZK astrophysical scenarios.

  These searches address the diffuse photon flux. The Pierre Auger Collaboration also performed photon searches from sources. In \cite{Aab:2014bha}, a blind search for point sources of EeV photons anywhere in the exposed sky  was performed. The search is sensitive to a declination band from $-85^\circ$ to $+20^\circ$, in an energy range from 10$^{17.3}$ eV to 10$^{18.5}$ eV. No photon point source has been detected with the consequence that no celestial direction exceeds 0.25 eV cm$^{-2}$ s$^{-1}$ in energy flux. To reduce the
  statistical penalty of many trials as done in the blind directional search, a targeted search from different source classes was pursued in \cite{Aab:2016bpi}. Several Galactic and extragalactic candidate
objects are grouped in classes and are
analyzed for a significant excess above the background expectation. No
evidence for photon emission from candidate sources is found, and upper limits are given for the selected candidate sources. These limits significantly constrain predictions of
EeV proton emission models from non-transient Galactic and nearby extragalactic sources. In fig. \ref{fig:PhotonLimits} right panel,  the particular case of the Galactic center region is illustrated.

\section{Photon background in the propagation of UHE cosmic rays}
\label{sec:PhBkgr}

The cosmic-ray energy spectrum and composition measured by UHECR experiments are strongly affected by
the propagation of particles from their sources to the Earth. Using propagation codes, it is possible to connect the injected spectrum/composition with the observed ones. Several investigations have been done in recent years to interpret UHECR data along this line\cite{Aloisio:2017ooo}. Most of these studies converge to scenarios with sources injecting hard spectra with
low rigidity cutoff and mixed composition, even though simplifying assumptions are used as uniform source distributions and 1D cosmic-ray propagation. All these results are strongly model dependent\cite{Batista:2015mea}: besides the hadronic interaction models which describe the shower development in the atmosphere, the other model uncertainties come from the photon background radiation which cosmic rays cross in their propagation and
the cross sections of photo-disintegration of nuclei interacting with background photons. These uncertainties are sizeable and mainly due to the lack of data\cite{Boncioli:2016lkt}.

\begin{figure*}[b!]
    \centering
    \begin{tabular}{ll}
      \includegraphics[scale=0.60]{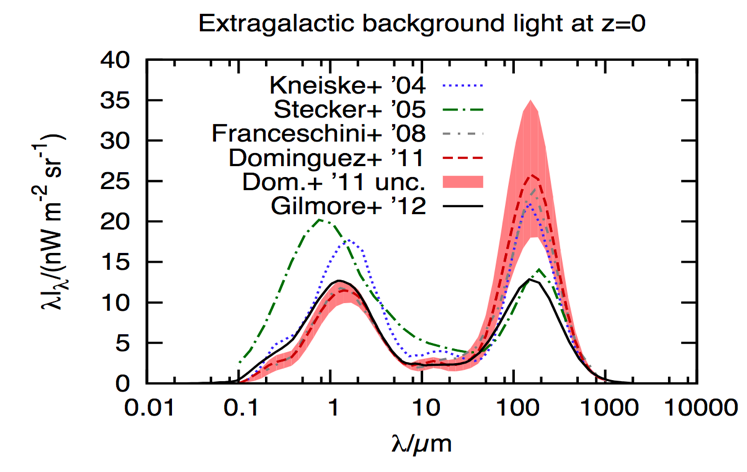}
      &
        \includegraphics[scale=0.30]{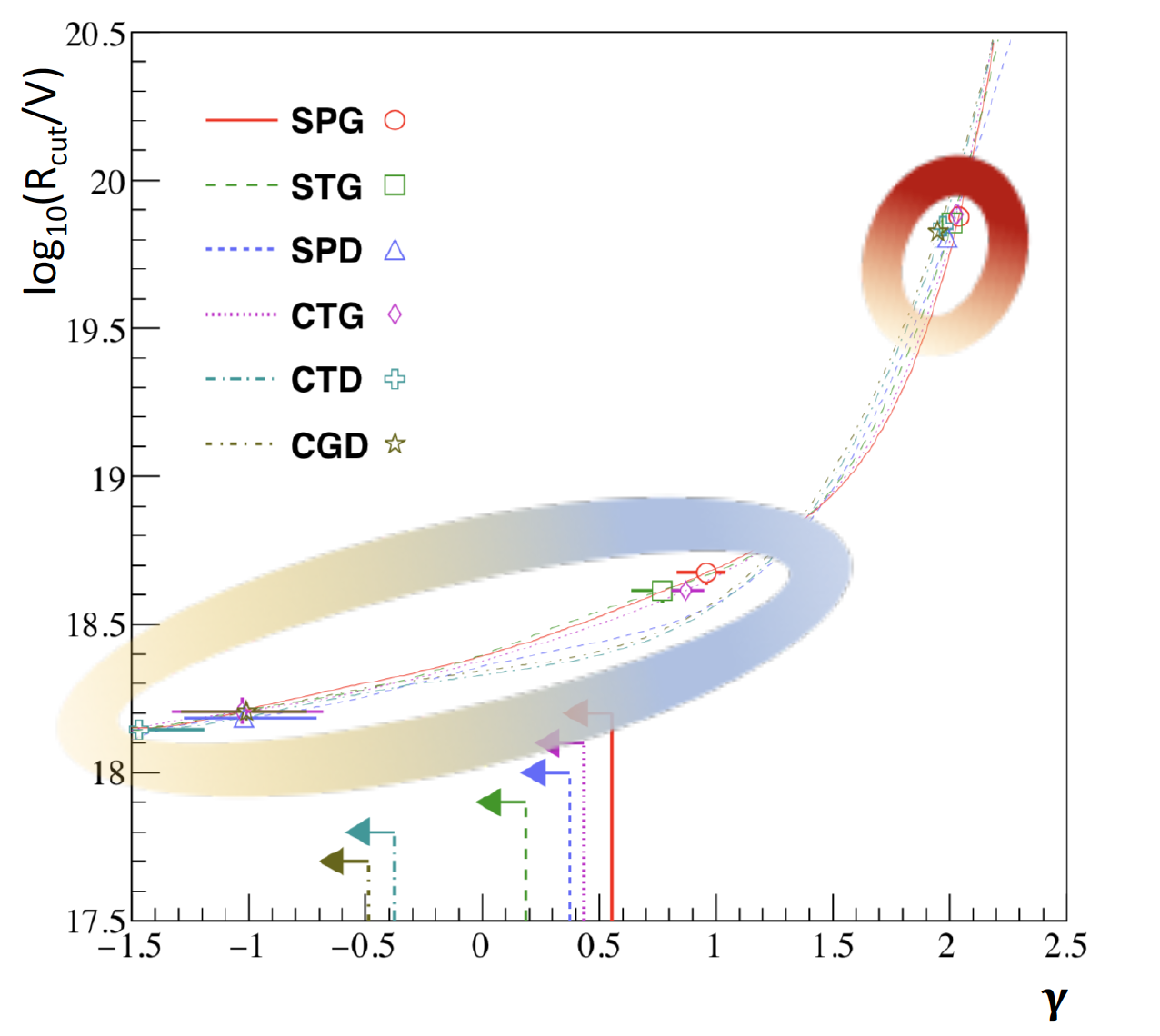}
        
    \end{tabular}
\caption{{\it Left: Intensity of the Extragalactic background light (EBL) at z= 0 for the models presented in ref.\cite{Batista:2015mea}. 
Right: The lines connecting the local minima for the six models given in ref.\cite{Aab:2016zth}. Symbols indicate the position of the minima of each model. Both the best fit at
$\gamma \lesssim 1$ (enclosed in the elongated ellipse at left) and the second local minimum at $\gamma \approx 2$ (in the small ellipse at right) are shown. }}
\label{fig:EBLeffects}
\end{figure*}
The Auger Collaboration has published a comprehensive study about the astrophysical implications from the combined fit of 
spectrum and composition data\cite{Aab:2016zth} above the ankle,  discussing in detail the effects of theoretical
uncertainties on the propagation of UHECRa and their interactions in the atmosphere as well as the dependence of the 
fit parameters on the experimental systematic uncertainties. In this study, we used a scenario
in which the sources of UHECRs are of extragalactic origin,
and nuclei are accelerated in electromagnetic processes with a rigidity-dependent maximum energy,
$R_\mathrm{cut}$. Within this scenario a good description of the shape of
the measured energy spectrum as well as the energy evolution of the $X_\mathrm{max}$ distributions can be
achieved if the sources accelerate a primary nuclear mix consisting of H, He, N and Si, if the
primary spectrum follows a power law $\propto E^{-\gamma}$ with a spectral index $\gamma \approx 1$ and if the maximum rigidity is about 10$^{18.7}$ V.
More details can be found in\cite{Aab:2016zth}.

Here I want to focus on the impact of the photon backgrounds on these outcomes.  In the energy range in which we are interested, the photon energy spectrum includes the cosmic microwave background radiation (CMB), and the infrared, optical and ultra-violet photons (commonly named extragalactic background light, EBL). The CMB has been measured extremely well, at least to the accuracy relevant for UHECR propagation. The EBL, which comprises the radiation produced in the Universe since the formation of the first stars, is relatively less known: several models of EBL have been proposed, among which there are sizeable differences (see fig. \ref{fig:EBLeffects} left panel),
especially in the far infrared and at high redshifts.

A quite general feature of the combined fit reported in\cite{Aab:2016zth} is a very definite correlation between the injection spectral index $\gamma$ and the rigidity cutoff $R_\mathrm{cut}$. Considering the deviance ($\approx \chi^2$) distribution there are in general two regions of local minima: one, which contains the best
minimum, corresponds to a low value of $R_\mathrm{cut}$ and $\gamma \lesssim 1$; a second relative minimum appears, less extended, around the
pair $\gamma \simeq 2$ and $\log_{10}(R_\mathrm{cut}/\mathrm{V}) \simeq 19.9$. In fig.
\ref{fig:EBLeffects} right panel, the positions of the minima of the combined
fit are shown for the six different models used in\cite{Aab:2016zth}; in particular, the last letter in the model names refers to the EBL model used (G = Gilmore+ '12,
D = Dominguez+ '11). One immediately notices the strong dependence of the best
solution on the EBL model. Instead, the region of the second minimum appears to have a modest dependence on all model parameters. To explain, at least partly, the different dependence on the photon background it is worth noticing that interactions on EBL photons become dominant as we approach low spectral indexes ($\gamma \lesssim 1$) and rigidity cutoffs ($\log_{10}(R_\mathrm{cut}/\mathrm{V}) \lesssim 18.7$). As a consequence, better models of EBL spectrum and evolution would help to reduce the uncertainties on the astrophysical scenarios.

\section{Photon fields in the CR source environments}
\label{sec:PhSources}
As shown in the previous section, a combined fit of spectrum and composition measurements allows one interpret Auger data above the ankle. An extension to lower energies is possible in two alternative secenarios. In the first one, the light component below the ankle originates from a different population of sources\cite{Aloisio:2013hya}. In this model, the spectrum injected by the sources of this component is steeper than the one corresponding to the other population. In the second scenario, the light component originates from the photo-disintegration of high energy and heavier nuclei in the photon field present in the environment of the source. This scenario has been proposed as a general mechanism in \cite{Unger:2015laa} and also in the context of the UHECR acceleration in more specific astrophysical objects\cite{Globus:2015xga,Fang:2017zjf,Biehl:2017zlw,Supanitsky:2018jje}.
It is worth pointing out that in this scenario we can also expect neutrinos emitted by the `extended source' (i.e., including the radiation region surrounding the UHECR accelerator) allowing in this way multimessenger studies to discriminate among the different astrophysical models.

\begin{figure*}[t!]
  \centering
  \includegraphics[scale=0.40]{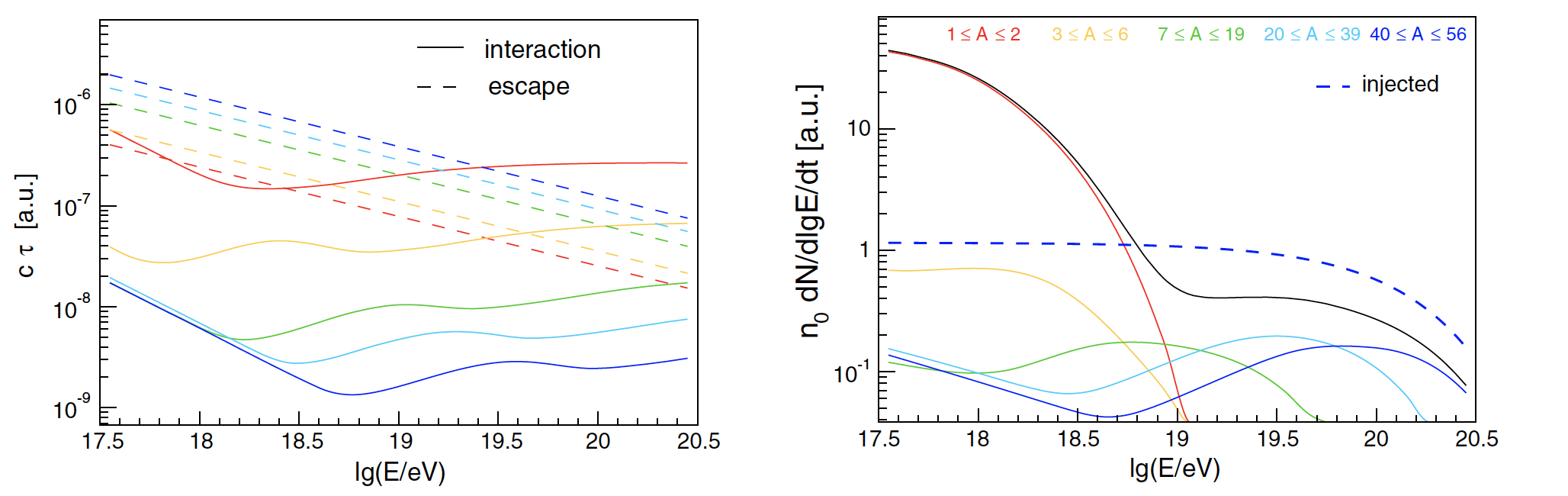}
\caption{{\it Interaction and escape times
for different nuclei (left). Case for injected $^{28}Si$ flux (dashed line) and
escaping fluxes (solid lines). Parameters are given in\cite{Unger:2015laa}. }}
\label{fig:UFA}
\end{figure*}

In ref. \cite{Unger:2015laa} it is shown that under certain hypotheses on the source parameters, the competition between interactions of nuclei emitted by the UHECR accelerator and the escape from the same region can generate {\it i.}~a spectrum feature consistent with the observed UHECR ankle, {\it ii.}~a mixed-composition escaping the source and {\it iii.}~protons
dominating in the ankle and sub-ankle regions. This fact is illustrated in fig.
\ref{fig:UFA} where $^{28}$Si nuclei are injected with a $E^{-1}$ spectrum in a
photon field represented by a broken power-law spectrum peaked at about 0.1 eV.

More studies are needed to understand if this promising scenario will give outcomes consistent with data in realistic astrophysical objects. In these studies a better description of the properties of the candidate source classes (luminosity and size of the accelerating region,
shape of photon spectrum and its peak energy) is mandatory to improve the comparison with the available data.


%
\end{document}